# Spatiotemporal Terahertz Emission Nanoscopy of Spintronic Photocurrents

F. Paries[*†1,2], R. Rouzegar[*3], J. Cai[4,5], M. Dai[4,5], F. Selz[1,2,6], J. Koelbel[6], G. Lezier[7], N. Tiercelin[7], D. Molter[2], D. M. Mittleman[6], G. von Freymann[1,2], X. Wu[†4,5,8], and T.S. Seifert[†3]

*equal contributions
[†]corresponding authors

[1] Department of Physics and Research Center OPTIMAS, RPTU Kaiserslautern-Landau, 67663 Kaiserslautern, Germany
[2] Fraunhofer Institute for Industrial Mathematics ITWM, 67663 Kaiserslautern, Germany
[3] Department of Physics, Freie Universität Berlin, 14195 Berlin, Germany
[4] Hangzhou International Innovation Institute, Beihang University, Hangzhou 311115, China
[5] School of Electronic and Information Engineering, Beihang University, Beijing, 100191, China
[6] School of Engineering, Brown University, Providence RI 02912 USA
[7] University of Lille, CNRS, Centrale Lille, Université Polytechnique Hauts-de-France, UMR 8520 - IEMN, 59000 Lille, France
[8] Zhangjiang Laboratory, 100 Haike Road, Shanghai 201204, China

**Abstract**

Capturing ultrafast spin and charge photocurrents on nanoscopic scales is essential for fundamental research in physics and engineering, as well as for future applications, such as novel spinorbitronic devices. Accessing the fundamental dynamics driven by changes in electronic energy, linear momentum, and angular momentum requires probing at its native spatiotemporal scales: femtoseconds and nanometers. However, experimental approaches achieving this simultaneous resolution remain scarce and instrumentally demanding.

Near-field probing offers promising platforms to combine ultrafast and nanometer resolution typically with high sensitivity to out-of-plane electric fields. However, applying this technique to in-plane ultrafast coupled spin and charge currents is largely unexplored, although being highly application-relevant – from ultrafast spin transport in 2D materials to spin-to-charge conversion in spintronic terahertz emitters (STEs).

Here, we fill this gap by performing spatiotemporal terahertz (THz) emission nanoscopy (TEN) of a photoexcited fiber-coupled STE using a scanning-probe microscope. We uncover a counterintuitive, dipolar spatial evolution of the near-field THz signal, which we show originates from the out-of-plane electric fields emerging from the in-plane spin-driven charge currents. Our findings explain why TEN is sensitive to ultrafast spin-driven in-plane charge currents, paving the way for TEN to become a fully vectorial probe for the spatiotemporal mapping of coupled nanoscale THz charge and spin dynamics.

## 1. Introduction

Gaining a detailed understanding of ultrafast charge and spin currents is vital for the development of next-generation technologies, from highly efficient solar cells and optimized light-driven chemical reactions to terahertz (THz) data processing devices[1-3].

In this context, the femtosecond (fs) time and nanometer (nm) length scales are of particular importance. These spatiotemporal scales not only match the dimensions of modern electronic and spinorbitronic devices that are envisioned to operate at ultrafast THz rates but also coincide with the native ranges that govern the relaxation of electronic systems in terms of energy, as well as linear and angular momentum[4]. Despite this, experimental approaches that combine such resolutions are scarce[5], often providing only indirect insights through experimentally demanding setups[6-8]. Consequently, there is an urgent need for probes capable of simultaneously capturing fs charge and spin transport at the nanoscale[9].

An established tool to measure ultrafast spin and charge photocurrents with fs time resolution is far-field THz-emission spectroscopy (TES)[10]. In TES, a fs optical excitation pulse triggers a charge-current burst, which subsequently radiates a THz electric field into the far field. By analyzing the polarization and amplitude of this radiation, TES can characterize the current in magnitude and direction, thus providing a vectorial probe of the ultrafast carrier dynamics. However, the spatial resolution of TES is fundamentally limited by diffraction to the micrometer range[11,12], preventing the visualization of local charge current distributions within or close to the excitation spot.

Recent advances have bypassed this diffraction limit by combining TES with near-field scanning probes, where a sharp metallic tip acts as a THz nano-antenna. In this THz-emission nanoscopy (TEN), the spatial resolution is limited only by the tip apex size – ideally reaching tens of nanometers[13]. This approach yielded remarkable local insights into ultrafast out-of-plane photocurrents in 2D materials[14], semiconductors[15-17], and metallic nanostructures[17-20]. However, a significant challenge remains: many fundamental ultrafast phenomena involve in-plane coupled electronic charge and spin dynamics. Because near-field probes are primarily sensitive to out-of-plane currents, it has remained unclear if and how TEN can provide spatiotemporal "filming" of these critical in-plane spin-driven charge dynamics on nanoscales.

Here, we use TEN to address this challenge by utilizing a fiber-coupled spintronic THz emitter (STE)[21-23] as a model system, where an ultrafast light-induced spin current is converted into a spatially confined in-plane charge-current distribution. By laterally translating the fiber-coupled STE beneath a stationary metallic tip of a scattering near-field optical microscope, we map the propagation dynamics in real space. Surprisingly, the detected THz electric field spatially evolves in a dipolar-like pattern. Supported by finite-element simulations, we demonstrate that this TEN signal originates from the out-of-plane electric-field components close to the STE surface that inherently accompanies in-plane charge currents in metallic films. These insights resolve the persisting ambiguity in the TEN community regarding the observation of ultrafast spin-related in-plane charge transport. In combination with the more established TEN-probing of THz out-of-plane currents, our findings pave the way for TEN to become a fully vectorial probe for the spatiotemporal imaging of ultrafast spin and charge dynamics on nanoscales applicable to a wider class of material systems.

## 2. Results

### 2.1 Fiber-coupled Spintronic THz-Emission Nanoscopy

To capture spin and charge currents with simultaneous fs and nm resolution, we integrate a fiber-tip spintronic terahertz emitter into a scattering near-field optical microscope (SNOM, *attocube*) (Fig. 1). This combination provides two important advantages: (i) the single-mode pump waveguiding guarantees a scan-position independent Gaussian excitation profile with sub-THz-wavelength dimensions, and (ii) the flexible-fiber coupling maintains a constant pump-probe delay and an invariant THz-detection response function throughout the scanning-probe motion.

*Fiber-coupled spintronic terahertz emitter.* Our localized THz source consists of an efficient tri-layer STE[21-24] that is sputter-deposited onto the facet of a single-mode glass fiber (Fig 1a-b). The STE is composed of a metallic thin-film heterostructure with a central ferromagnetic layer (CoFeB, 1.8 nm thickness) and two adjacent heavy-metal layers with large but opposite spin-Hall angles (W and Pt, 2 nm thickness each). For THz generation from the STE, a femtosecond optical pulse incident from the fiber (Fig. 1c) excites the CoFeB layer triggering ultrafast demagnetization and a concurrent spin-current injection into the heavy metal layers[24,25]. Due to their large and opposite spin-Hall angles, these spin currents are converted into a transverse in-plane charge-current burst via the inverse spin-Hall effect (ISHE) inside the heavy metals, radiating a THz electromagnetic pulse (Fig. 1a).

*Nanoscale spatiotemporal mapping.* To perform non-local nanoscale imaging, the fiber-coupled STE is mounted to the SNOM's 3-axis piezo stage and positioned underneath the sharp SNOM tip (Fig. 1c). The metallic SNOM tip is primarily sensitive to out-of-plane electric fields ($E_z$), with its nanometer-scale apex locally enhancing near-fields by over an order of magnitude (see *Supplementary information*). By oscillating the SNOM tip above the sample at a frequency $\Omega$, we extract the near-field contribution via demodulating the signal at higher harmonics $n\Omega$ (Ref. [13,26]). The emitted THz electric field is collimated by an off-axis parabolic mirror and focused onto a photoconductive antenna allowing for time-domain sampling of the THz waveform (Fig. 1d, see *Materials and methods*).

Importantly, because the fiber-coupled path eliminates changes in the optical path as it bends slightly throughout raster-scan movements, we can directly visualize the spatiotemporal evolution of the THz dynamics with a constant THz-detection response function at every scan position of the SNOM.

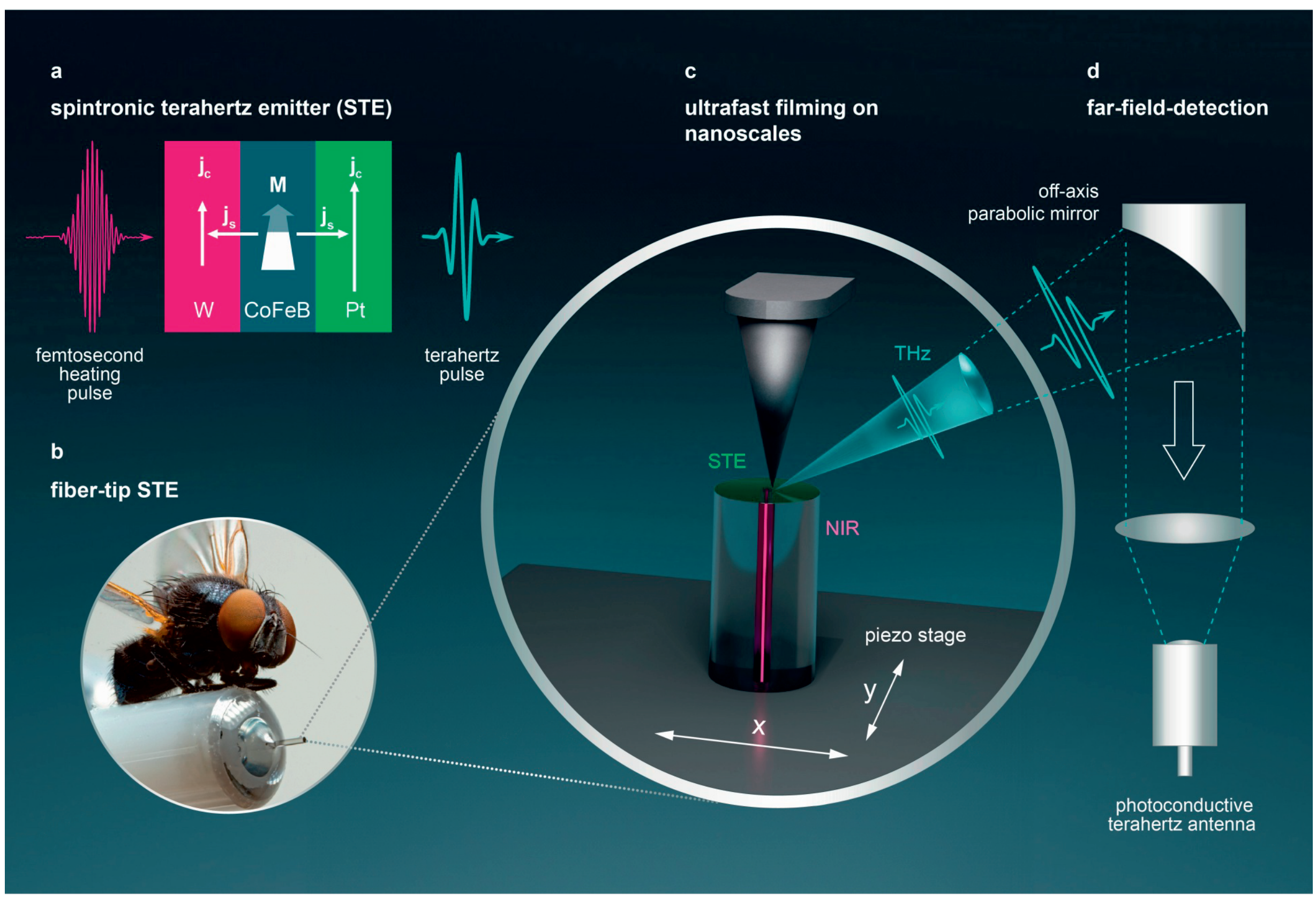

**Fig. 1 | Ultrafast filming on nanoscales. a** Working principle of a spintronic terahertz emitter (STE) based on a W|CoFeB|Pt trilayer. A femtosecond laser pulse excites the metallic heterostructure, generating a transient excess of spin in the ferromagnet that injects spin currents into the adjacent heavy metals. These spin currents are converted into in-plane charge currents via the inverse spin Hall effect, leading to constructive addition of the two charge currents due to the opposite spin-Hall angles of W and Pt. **b** Fiber-tip STE fabricated by sputtering the trilayer directly onto the tip of a single-mode fiber. The fly provides an intuitive sense of scale. **c** Spatiotemporal THz-emission nanoscopy enabled by laterally translating the fiber-coupled STE under the metallic tip of a scanning-probe microscope. The tip is sensitive to the local out-of-plane component of the THz electric near-fields above the STE. **d** The emitted THz electric field is focused onto a photoconductive antenna that is transiently becoming conductive through excitation by a probe pulse, leading to an induced photocurrent proportional to the THz-field amplitude. A full THz waveform is sampled by varying the pump-probe delay.

## 2.2 Experimental Results

*THz-emission nanoscopy movies from the fiber-coupled STE.* We obtain THz-emission movies with nanometer spatial resolution by tracking the THz-emission signal as a function of the fiber-coupled-STE position beneath the SNOM tip (see *Materials and methods*). We record 3D datasets with 1 temporal and 2 spatial dimensions in a pump-probe fashion (see doi.org/10.5281/zenodo.18739933). Due to the fully fiber-coupled pump path, the temporal delay axes are identical for each pixel, eliminating any relative shift. Figure 2a shows six first-harmonic movie frames of the THz-emission signal, lock-in detected at the tip-oscillation frequency, for distinct detection-time delays. Intriguingly, the spatial pattern follows a clear dipolar pattern. This pattern is centered at the fiber core, which reveals a surprising result: no THz-emission signal is detected at the center of the Gaussian STE-excitation spot where the pump intensity and, thus, the current density is maximum. Instead, the off-centered dipole minima and maxima oscillate over time, with the dipole axis oriented perpendicularly to the in-plane STE magnetization ($\mathbf{M}$). This dipole-to-magnetization orientation has also been confirmed in a previous spintronic TEN study[20].

To understand these dipole features, we first recall that the SNOM metallic tip selectively enhances the out-of-plane electric-field component ($E_z$). We combine this with the analytical solution for the electric-field distribution surrounding a point-like in-plane electric dipole embedded in a thin metallic film extending to infinity in the plane[27]. The pattern of the out-of-plane electric field $E_z$ for a delta-like in-plane dipole density follows:

$$E_z \propto \frac{1}{r^2} \cos \Phi. \tag{1}$$

Here, $r$ is the distance from the dipole center and $\Phi$ is the angle in the sample plane with respect to the dipole axis. Equation 1 describes a two-lobe structure with a center zero line that decays according to $1/r^2$, qualitatively matching our experimental results (Fig. 2). Consequently, we conclude that the spintronic TEN images are the $E_z$ signature of the in-plane charge-current density driven by the laser-excited STE.

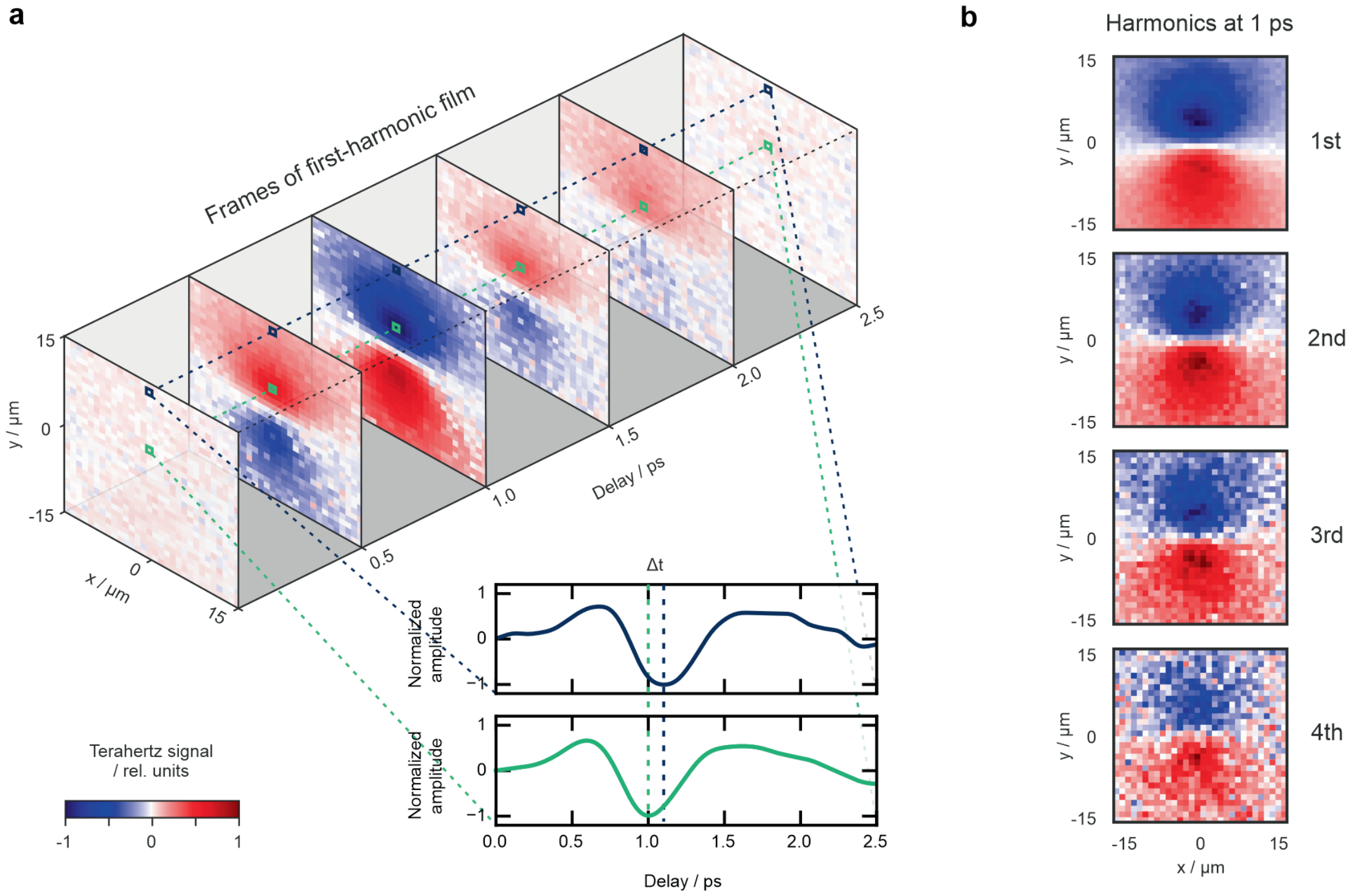

**Fig. 2 | Time-domain measurements. a** Selected frames of the recorded laser-induced in-plane terahertz dynamics. An area of 30×30 µm$^2$ is raster-scanned with a step size of 1 µm and an acquisition time of 84 s per pixel. For each pixel a THz waveform is acquired by varying the pump-probe delay with a mechanical delay line. Due to the fully fiber-coupled pump path, the pump-probe delay is immune against any raster-scan movement, and the THz signal resulting from ultrafast in-plane dynamics can be filmed. Two selected waveforms showcase the propagation delay of the THz charge current along the surface. **b** THz-emission signals at a fixed delay of 1 ps obtained by demodulating the detected THz signals at different harmonics of the tip-tapping frequency. All harmonics (1st – 4th) exhibit an identical spatial dipole profile yet with decreasing signal-to-noise ratios.

Based on an inverse-spin-Hall-effect driven spin-to-charge current conversion[22], the transient charge-current dipole is expected to be oriented orthogonally to the STE's in-plane magnetization $\mathbf{M}$ (Fig. 1a). In our measurement (Fig. 2), with $\mathbf{M}$ parallel to the x-axis, the observed dipole axis is indeed orthogonal to the direction of $\mathbf{M}$. This agreement bolsters our interpretation of the spintronic TEN image contrast and aligns with related SNOM studies of in-plane electric dipoles[28-30].

Furthermore, this mechanism also explains why previous TEN studies[19,20] could obtain THz-emission signals from STEs despite the supposed lack of in-plane sensitivity: in those works, where femtosecond excitation is realized via free-space sample excitation, the THz-emission signal was likely optimized by unintentionally moving the excitation-laser spot relative to the SNOM tip. This created an off-centered pump-tip configuration, allowing the tip to pick up the finite out-of-plane electric field components at the edges of the in-plane charge-current distribution.

*Temporal evolution and propagation.* We continue our analysis by extracting the THz signal for a given scanning position. A clear oscillating waveform with subpicosecond dynamics vs. detection-time delay emerges typical of STEs. Remarkably, a propagation delay becomes visible when comparing THz waveforms at locations 4 and 14 µm away from the emitter center (Fig. 2a). We tentatively assign this to the velocity of the surface plasmon polariton moving along the air/metal interface[31]. This propagating nature is even more apparent in the full spintronic TEN movies (see doi.org/10.5281/zenodo.18495495) and will be analyzed in more detail later.

*Higher-harmonic analysis.* As shown in Fig. 2b, the spintronic TEN images recorded with higher-harmonic demodulation ($1^{\text{st}}$ to $4^{\text{th}}$) follow identical spatial patterns with decreasing signal-to-noise ratios. While higher-harmonic demodulation typically increases the spatial resolution by restricting the interaction window to the moment the SNOM tip is closest to the STE[13], our results suggest we are already resolving the dominant spatial features at the first harmonic with our 1-µm experimental scan step size. The spintronic TEN images are dominated by the long-range $1/r^2$ decay (see Eq. (1)). Notably, nanometer features, such as the polarity flip at the dipole center, can be still resolved.

*Frequency-domain analysis.* After concluding our time-domain analysis, we now turn our attention to the frequency domain. Figure 3a shows the spectral amplitude and spectral phase obtained by Fourier transforming the time-domain signal at the spatial location with the largest peak-to-peak amplitude (see gray arrow). The obtained center frequency of approximately $\omega/2\pi = 1\ \mathrm{THz}$ is consistent with the laser pulse duration of 70 fs and the THz-detector sensitivity. The spectral phase is relatively flat indicating a minor chirp of the THz waveform as typical for STEs[22].

The spectral-amplitude map at a frequency of 1 THz (Fig. 3b) reinforces the dipolar spatial profile observed in the time domain (Fig. 2). A slight asymmetry along the y-axis is evident, likely originating from a minor change in tip-sample distance during the scan. In the spectral-phase map (Fig. 3c), we observe an abrupt, nanoscale $\pi$ phase jump at $y = 0$, as predicted by the sign change of the $\cos\Phi$ in Eq. 1. Additionally, the spectral phase exhibits a constant radial gradient that switches sign at $y = 0$. This phase gradient is the frequency-domain signature of a propagation delay observed in the time domain (Fig. 2a).

Note that to increase the signal-to-noise ratio, all images obtained for different harmonic demodulations shown in Fig. 2b are summed up prior to Fourier transformation. Minor diagonal stripes in the phase map in Fig. 3c arise from periodic instabilities of the SNOM during the extended measurement duration (see *Materials and methods*).

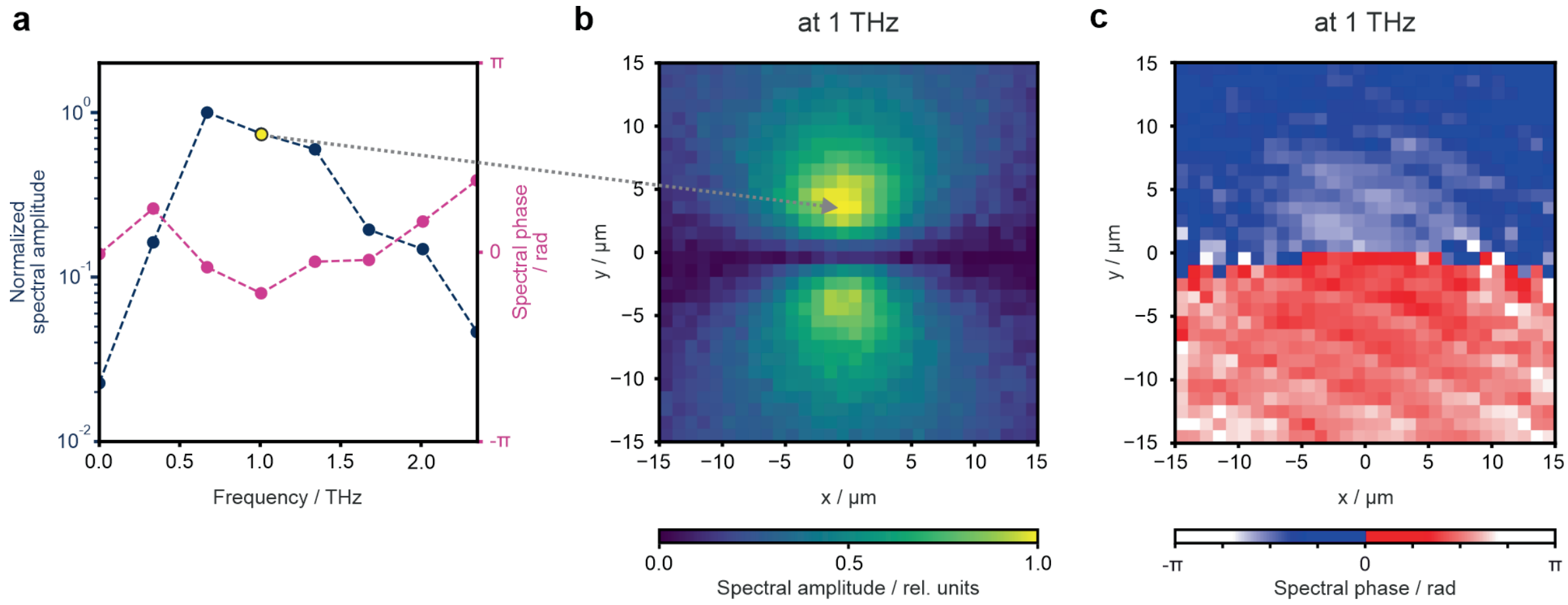

**Fig. 3 | Frequency-domain analysis. a** Spectral amplitude and spectral phase at the spatial position of maximum THz signal. A typical terahertz spectrum is measured with a frequency resolution of 0.33 THz, limited by the delay range of 3 ps. **b** Map of the spectral amplitude at 1 THz relative to the maximum. A clear spatial dipole profile is visible without any signal on the center line along the direction of the sample magnetization.

**c** Map of the spectral phase at 1 THz. The spectral-phase map exhibits a radial gradient indicative of spatial terahertz-current dynamics, i.e., propagation delays along the surface.

### 2.3 Finite-element Simulations

*Simulating the spintronic TEN experiment.* While Equation (1) provides a qualitative understanding of the observed spintronic TEN image contrast under idealized conditions, a rigorous description requires detailed modelling of the experimental geometry. We account for the finite size of the fiber-coupled STE and the complex photonic environment using a finite element solver (COMSOL Multiphysics, RF module). We simulate the experimental geometry, including the fiber facet, thin metallic layer with a 6 nm thickness and scanning probe, driven by a Gaussian-shaped in-plane electric charge-current density, reflecting the fiber-mode symmetry. From this, we extract the out-of-plane electric field component $E_z$ at 1 THz (see *Materials and methods*).

*Comparison between simulation and experiment.* In Fig. 4, we present a back-to-back comparison of the simulation results and experimental measurements. The THz-spectral-amplitude maps at 1 THz for scan step sizes of 1 µm and 300 nm show remarkably good agreement with the simulation (Fig. 4a). This consistency further substantiates our interpretation: the TEN signal arises from the local $E_z$ components generated by a spatially confined in-plane current source.

The agreement is further exemplified by comparing line scans at $x = 0\ \mu\text{m}$ (Fig. 4b), which reveal an excellent match between the measured and simulated peak positions, as well as the center-dip shape with a $1/e^2$ width of approximately 400 nm. We note that the measured line-cut exhibits a slightly shallower spatial falloff than the simulation. This broadening likely indicates a deviation from a purely Gaussian current profile. In our simulation, we assume a Gaussian profile in line with the fiber mode profile; however STEs can exhibit saturation at incident pump fluences exceeding 0.5 mJ/cm$^2$ (Ref. [32]). Given our experimental conditions, saturation at the center of the excitation spot likely results in a "flat-top" current profile, leading to the observed spatial broadening.

We also extract the spectral phase along the line cut at $x = 0\ \mu\text{m}$ (Fig. 4c), which follows a clear linear trend in both experiment and simulation. Linear fitting of the simulation data yields a phase gradient of $\pm 0.011\pi\ \mu\text{m}^{-1}$, corresponding to a subluminal 1-THz phase velocity of about 60% of the speed of light. Such subluminal velocities have been reported for similar surface plasmon-polariton modes[33].

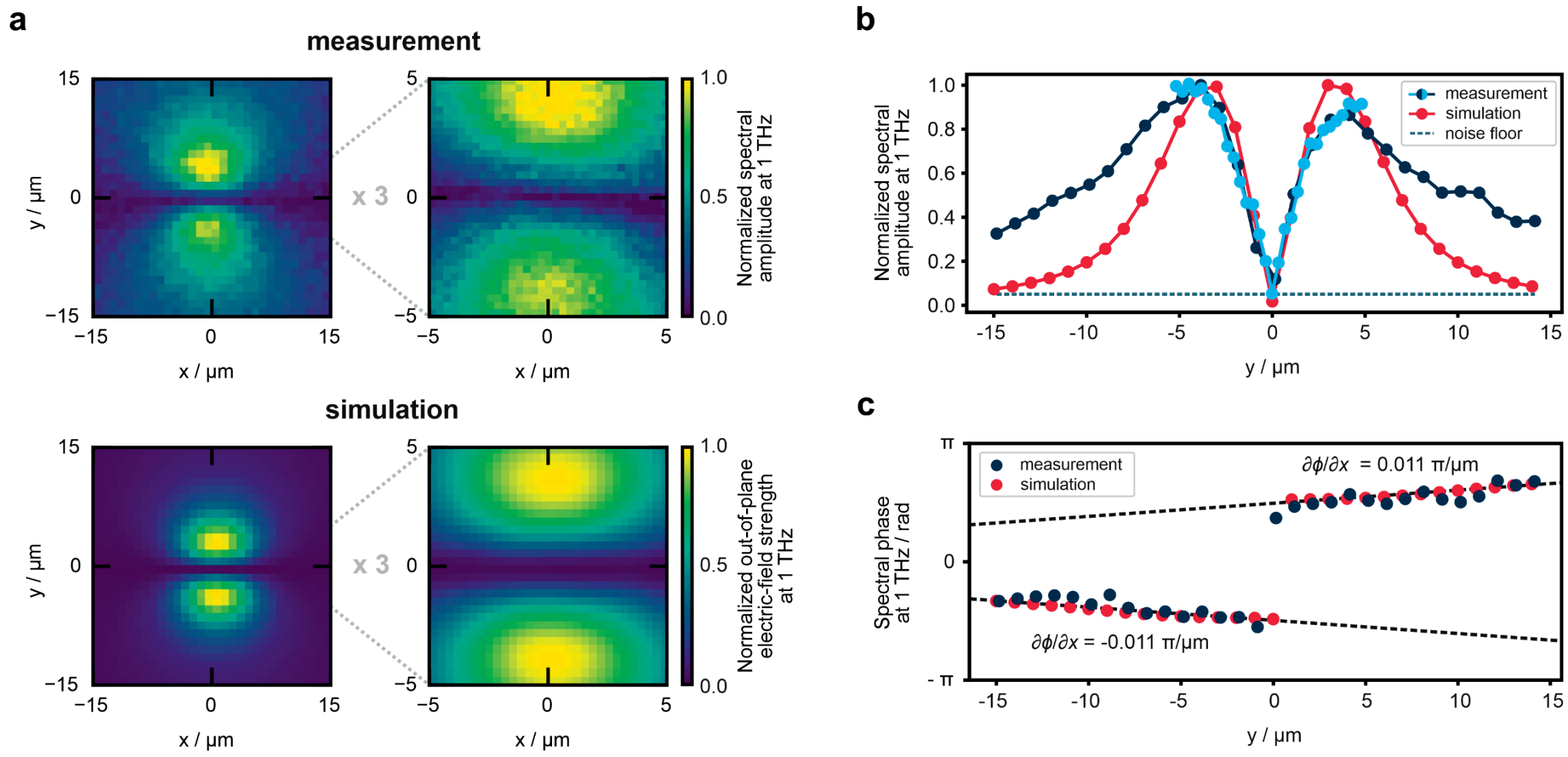

**Fig. 4 | Measurement vs. simulation at 1 THz. a** Maps of the spectral amplitudes at 1 THz for scan step sizes of 1 µm (left) and 300 nm (right). The simulations display the out-of-plane electric-field component $E_z\ (\omega, x, z)$ at $\omega/2\pi$=1 THz. **b** Center line-scans at $x = 0\ \mu\mathrm{m}$ from the data shown in panel a along with simulation results. **c** Spectral phases for the line scans shown in b along with the phase gradient obtained by a linear fit of the simulation data.

*Optimizing experimental conditions.* To explore the robustness of the presented technique, we performed a series of simulations varying the spatial extent of the in-plane-current distribution (Fig. 5a) and the average metal-film conductivity (Fig. 5b). In these simulations, the initial peak charge-current density $j_0$ is fixed at 1 A/m$^2$ (see *Materials and methods*).

Importantly, our single-mode fiber-tip STE design, with a pump mode-field diameter (MFD) of 10.5 µm, is near-optimal for efficient $E_z$ generation. For MFD values exceeding 25 µm, the finite geometry limited by the fiber diameter of 125 µm begins to interfere with the spatial pattern of $E_z$, as the fields can no longer be considered decaying in an infinite plane. This confirms that Eq. 1 remains valid only when the current source is much smaller than the underlying geometry and far away from physical boundaries.

Interestingly, we find that the average STE metal-film conductivity primarily scales the amplitude of $E_z$ without altering its spatial distribution (Fig. 5b). For conductivities exceeding $10^4$ S/m, the $E_z$ amplitude starts saturating, making the TEN signal robust against moderate conductivity fluctuations in the experimentally relevant region around $10^6$ S/m (Ref. 22,34). This stability simplifies the interpretation of spintronic TEN images significantly, as spatial variations in film quality do not require complex correction factors. Conversely, it also highlights that the TEN signal strength will decrease significantly when investigating nonmetallic or poorly conducting samples.

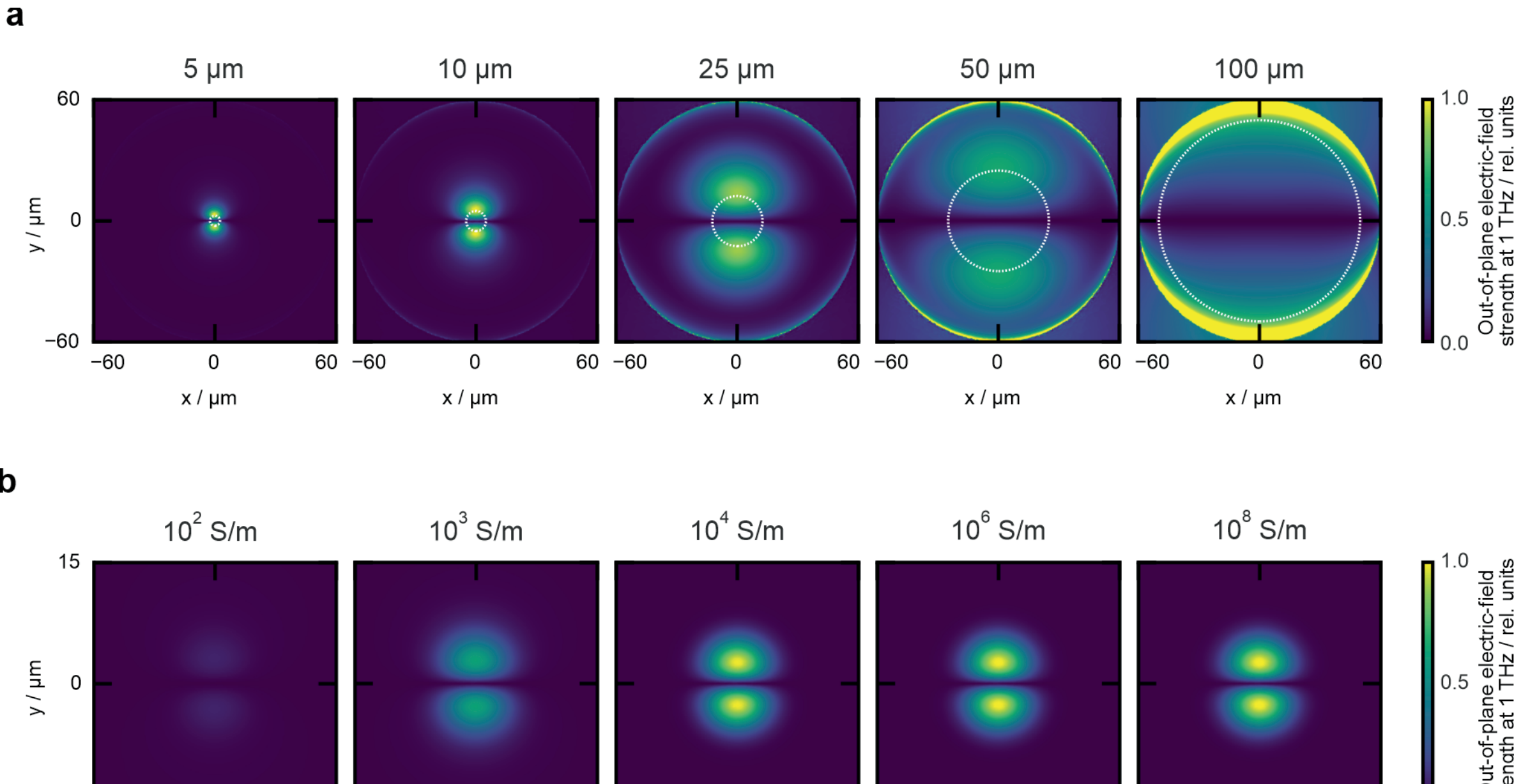

**Fig. 5 | Simulated maps of the out-of-plane electric field $E_z\ (\omega, x, z)$. a** Dependence of $E_z$ on the pump-mode-field diameter, given by the value indicated above each panel and by the white dotted circles. **b** Dependence of $E_z$ on the average STE metal-film conductivity. Note that in the experiment, the mode-field diameter is $10.5\ \mu m$ and the average STE metal-film conductivity is $10^6\ S/m$, which is near-optimal for efficient out-of-plane electric-field generation. In all simulations, the primary current density $j_0$ is fixed to a peak of $1\ A/m^2$ implying that the pump power scales quadratically with the mode-field diameter.

## 3. Discussion

To contextualize our results, we note that our fiber-coupled TEN system offers significant advantages over state-of-the-art TEN approaches. In standard, non-fiber-coupled SNOM setups, one typically encounters a trade-off: when the tip is moved while keeping the sample and excitation spot stationary, the setup-response function changes severely as the tip position varies with respect to the THz-detection unit[35]. Alternatively, when laterally translating the sample under a stationary tip, the setup-response function remains constant, but instead changes in optical pump-path length, excitation-profile and -position on the sample complicate image interpretation. Conversely, our fiber-coupled setup ensures invariance for all three during scanning: invariant excitation-profile and -position, constant THz-detection delay, and a constant setup-response function. This enables real-space visualization of out-of-plane and in-plane currents with simultaneous femtosecond and nanometer resolution.

In conclusion, we film the in-plane spin-driven charge dynamics in a fiber-coupled spintronic terahertz emitter in real space. Our simulations reproduce the measured near-field spatiotemporal maps providing access to the fundamental image-forming mechanism behind spintronic terahertz-emission nanoscopy: the measured signal originates from the out-of-plane electric-field component that inevitably accompanies localized in-plane current distributions. This resolves an ambiguity in the TEN community and explains why previous studies could detect in-plane currents despite the probe's inherent vertical sensitivity.

We foresee broad applications of this table-top technique in heterogeneous material systems and nanostructures, including the study of magnetic nanodomains in altermagnets[39] or 2D magnets[40]. Beyond magnetism, TEN is now equipped to study photocurrent heterogeneities in a wide range of systems on their native spatiotemporal scales.

## Materials and methods

### *Details of measurements*

For the measurements visualized in Fig. 2 and Fig. 3, an area of 30×30 μm² is raster-scanned by moving the fiber-coupled STE with a step size of 1 μm. Per pixel, three THz waveforms are recorded, each sampled in 85-fs steps over a 3-ps delay window, with a lock-in time constant of 800 ms per step. These three waveforms are then averaged to yield a single waveform per pixel acquired over a total acquisition time of 84 s. To achieve higher spatial resolution (see zoom in Fig. 4), the raster-scan area is reduced to 5×5 μm² and the measurement is repeated with a step size of 300 nm.

The fiber-tip STE is pumped by an Erbium-doped fiber-laser system with a repetition rate of 100 MHz (*Menlo ELMO* and *Menlo ELMA*). The average pump power is set to 2 mW using a variable optical attenuator positioned just before the fiber-tip STE. The total pump-fiber length is chosen such that the pump-pulse duration reaches its Fourier limit at the fiber-tip STE (70 fs FWHM). The 1/e² mode-field diameter of the used single-mode pump fiber is specified as 10.5 μm. A detailed illustration of the experimental setup is shown in Fig. S1.

### *Correction of periodic instrument drifts*

Over the measurement duration of approximately 21 hours, a periodic drift of the instrument was evident across several independent signal channels. It was most pronounced in the z-axis channel of the piezo stage and likely caused by temperature fluctuations within the laboratory. These instabilities caused the acquired waveforms to periodically drift in temporal delay (see Fig. S5). To correct this drift, a sinusoidal two-dimensional delay-correction map parametrized by a single spatial frequency, phase, and amplitude is defined. The frequency and phase are obtained by applying a spatial Fourier transform to the two-dimensional map of the scalar z channel (z-axis of the piezo stage) and selecting the spectral-peak value. The amplitude is set to 56 fs, which minimizes the spatial distortion of the THz signal at a fixed delay. The delay-correction map then defines a delay shift for each pixel that is subsequently applied to the measurement data.

### *Details of finite-elements simulation*

We use COMSOL Multiphysics to simulate THz emission from a fiber-tip spintronic THz emitter (STE). The primary charge current is assumed to be proportional to the absorbed pump energy density. To isolate purely electromagnetic effects and exclude the spintronic contribution in the simulations, we prescribe a primary charge current inside a homogeneous metallic film[34] (thickness $d = 6\,\mathrm{nm}$ and conductivity $\sigma = 1 \times 10^6\ \mathrm{S/m}$) as $j_{\mathrm{primary}}(r) = j_0 \exp(-2r^2/w)$ where $r$ is the radius, $w$ is beam waist (with $\mathrm{MFD} = 2w$), and $j_0$ is the amplitude of the primary current density. In all simulations presented in this work, we fix $j_0 = 1\ \mathrm{A/m^2}$, independent of mode-field diameter, film thickness, and electrical conductivity.

We map both the in-plane and out-of-plane electric fields and charge-current distributions, as shown in the *Results* and the *Supplementary information*. We observe that the in-plane electric field is continuous across the film–air interface, whereas the out-of-plane component vanishes inside the metallic film and exhibits a discontinuity at the interface, becoming nonzero in the adjacent air region. Accordingly, all out-of-plane electric-field results presented in this paper correspond to the field evaluated immediately outside the metallic interface in air.

## Supplementary information

### *Experimental setup*

As depicted in Fig. S1, the output of an Erbium-doped femtosecond laser (100 MHz repetition rate) is split into two trains of pump and probe pulses, each amplified by an Erbium-doped fiber amplifier (EDFA) to an average power of 20 mW. To stay below the damage threshold of the single-mode fiber-tip[32], the pump arm is attenuated to 2 mW using a variable optical attenuator. The fiber-tip STE is integrated into a commercial THz SNOM system (*attocube*) using a custom holder mounted on the system's three-axis piezo stage. The nanoscopic THz SNOM tip emits a terahertz field into the optical far field that is modulated at the tip's 48 kHz tapping frequency. A parabolic mirror collimates the radiation before it is focused by a Silicon lens onto a photoconductive antenna, that becomes transiently conductive upon probe-pulse excitation. The THz-induced photocurrent is transimpedance amplified and the resulting signal lock-in detected. By imposing a varying time delay between the THz and the probe pulse, the terahertz waveform is sampled in the time domain. Notably, the THz signal bandwidth is not limited by the PCA detector but rather by the signal-to-noise ratio. THz SNOM setups inherently favor lower terahertz frequencies, since the tip and the tip-sample junction act as a low-pass filter[9,41].

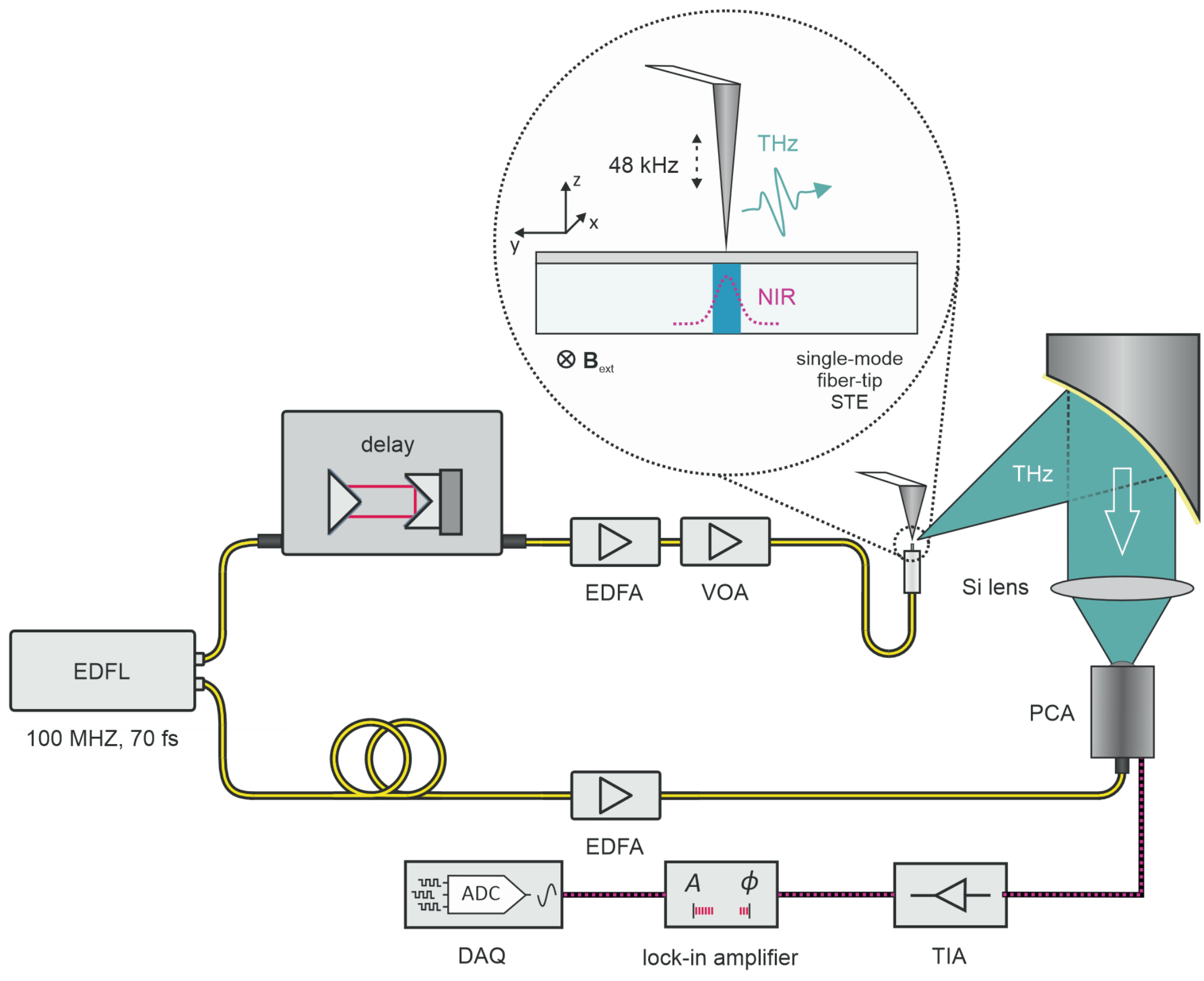

**Fig. S1 | Illustration of the experimental setup.** A single-mode fiber-based optical sampling setup is combined with a fiber-tip spintronic terahertz emitter and a THz SNOM system. EDFL – Erbium-doped fiber laser, EDFA – Erbium-doped fiber amplifier, VOA – variable optical attenuator, PCA – photoconductive antenna, TIA – transimpedance amplifier, DAQ – data acquisition.

### *Simulated charge-current densities and electric-field components*

Upon absorption of the femtosecond pulse, a transient spin current is injected from the ferromagnetic layer into the adjacent heavy-metal layers and converted into a primary in-plane charge current perpendicular to the magnetization direction of the ferromagnetic layer. This primary in-plane charge-current density $\mathbf{j}_{\text{primary}}$ follows the spatial intensity profile of the gaussian-shaped pump-field. Due to the layer's metallic nature with a conductivity of approximately $10^6$ S/m, the sheet conductance of the spintronic THz emitter is sufficiently high to allow for a significant induced currents despite the small thickness of 6 nm according to Ohm's law. The sum of the primary charge-current density $\mathbf{j}_{\text{primary}}$ and this induced charge-current density $\mathbf{j}_{\text{ind}}$ is visualized in Fig. S2a.
Since the size of the primary Gaussian charge-current-density profile, and thus the scale over which the in-plane electric-field components $E_\text{x}$ and $E_\text{y}$ spatially vary, is comparable to or smaller than the THz wavelength inside the material, a significant out-of-plane electric field component emerges that can couple efficiently to a metallic SNOM tip (Fig. S2b).

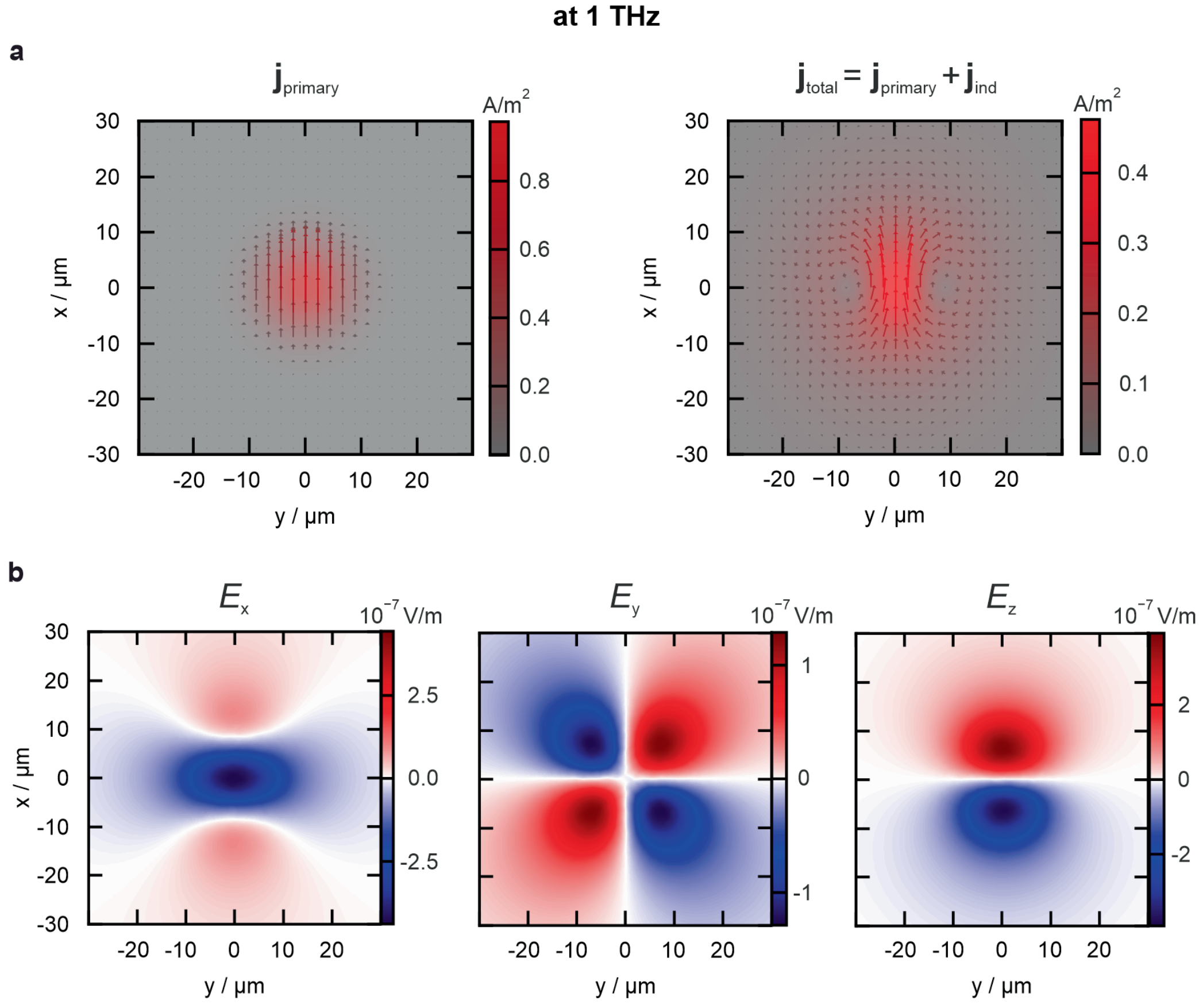

**Fig. S2 | Simulated charge-current densities and electric-field components. a** Vector-field representation of the simulated primary charge-current densities $\boldsymbol{j}_{primary}$ and simulated total charge-current densities $\boldsymbol{j}_{total} = \boldsymbol{j}_{primary} + \boldsymbol{j}_{ind}$ , where $\boldsymbol{j}_{primary}$ maps the gaussian pump-mode field profile and $\boldsymbol{j}_{ind}$ is the induced charge current governed by the material properties. Turning on the metallic thin-film properties ($\boldsymbol{j}_{primary} + \boldsymbol{j}_{ind}$), two vortices emerge. The color mapping visualizes the absolute value of the vector field. **b** Simulated electric-field distributions related to the current densities in a.

### *Electric-field enhancement*

We calculate the effect of field enhancement of the tip by simulating an STE with a tip close to the metallic film. A metallic tip with an apex of 100 nm and conductivity of $\sigma = 1.2 \times 10^7$ is considered. We observe that $E_z$ is enhanced by about an order of magnitude within the vicinity of the metallic nanoscopic tip as showcased in the simulated electric field in Fig. S3. This causes a charge-current density to be induced within the tip that is similar in magnitude as the primary in-plane current inside the STE.

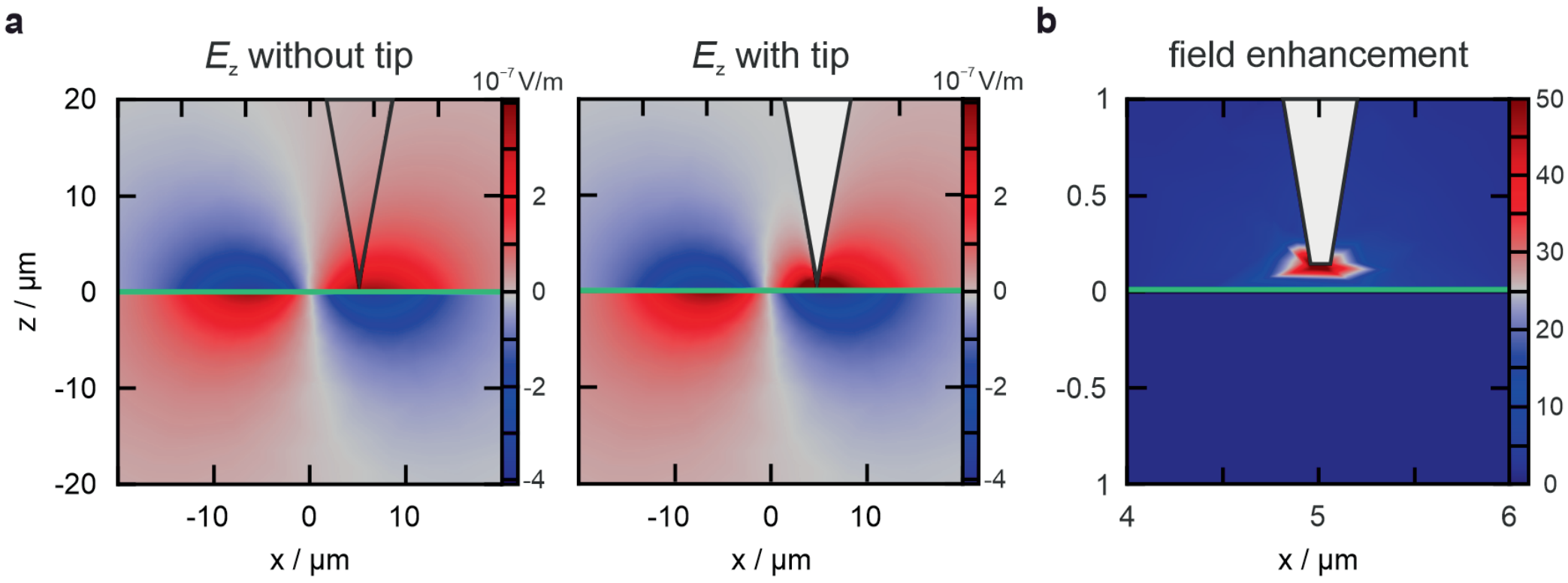

**Fig. S3 | a** Simulated $E_z$ without and with the presence of a sharp metallic tip. The STE is illustrated in green. **b** Visualization of the tip-induced light-field enhancement. The out-of-plane electric field is enhanced by more than one order of magnitude. The edgy appearance is caused by the limited resolution of the simulation mesh.

## *Conductance dependence*

In this section, we investigate the effect of film thickness on the $E_z$ component responsible for the measured THz signal. In practice, increasing the film thickness reduces the absorbed pump energy density. However, since our goal is to isolate the dependence of $E_z$ on film thickness, we normalize all simulations to the same absorbed energy density. Equivalently, the excitation strength (and thus $j_0$) is kept identical for all thicknesses.

We fix the conductivity at $\sigma = 10^3$ S/m and vary the film thickness between $d = 5$, 10, and 20 nm. The simulations show that $E_z$ increases with increasing thickness. When this result is combined with Fig. 5b of the main manuscript, which shows that $E_z$ increases with $\sigma$, we conclude that $E_z$ scales with the film conductance ($G = \sigma d$) rather than with the bulk conductivity alone.

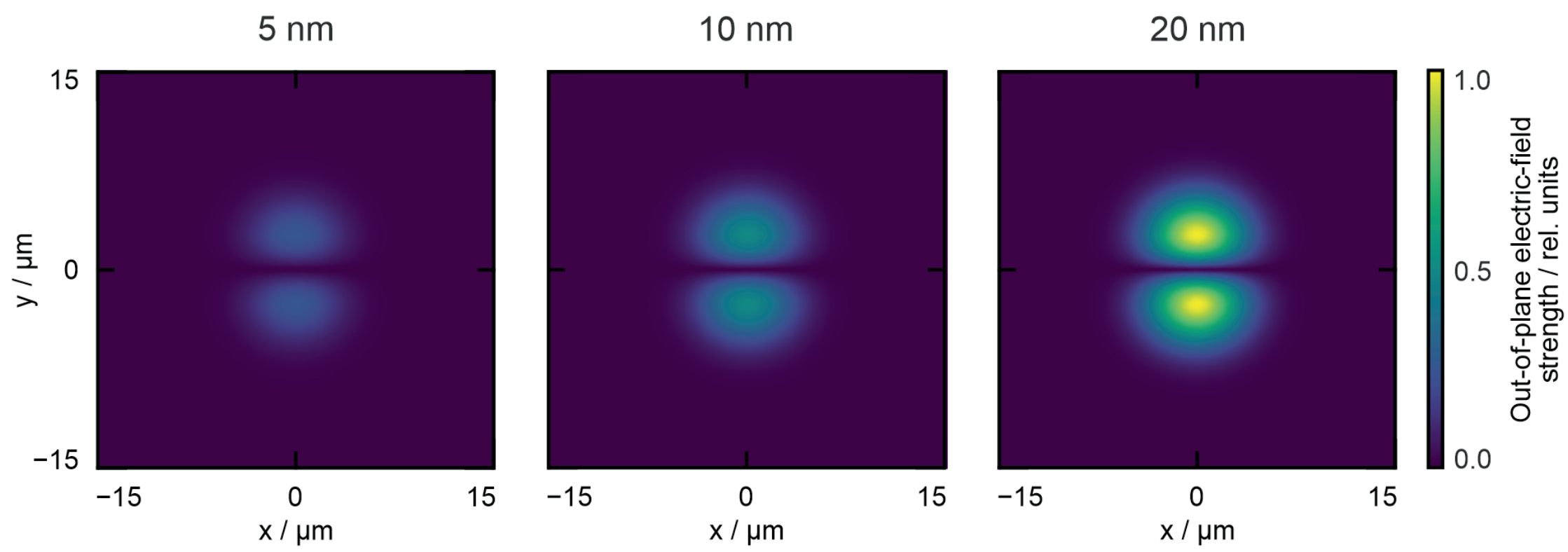

**Fig. S4 | Simulated maps of the out-of-plane electric field $E_z$ $(\omega, x, z)$.** Dependence of $E_z$ on the film thickness while keeping conductivity $\sigma = 10^3$ and energy density (or equivalently same primary current) same for all thicknesses

*Correction of periodic instrument drifts*

Fig. S5 illustrates the delay-correction procedure described in *Materials and methods*. Repetitive temperature variations on the time scale of hours result in thermal expansions and contractions that cause the piezo-stage control loop to adjust the z-axis position during the measurement to keep the average tip-sample distance constant. The z-channel readout thus provides a calibration signal to correct for the resulting periodic delay shifts.

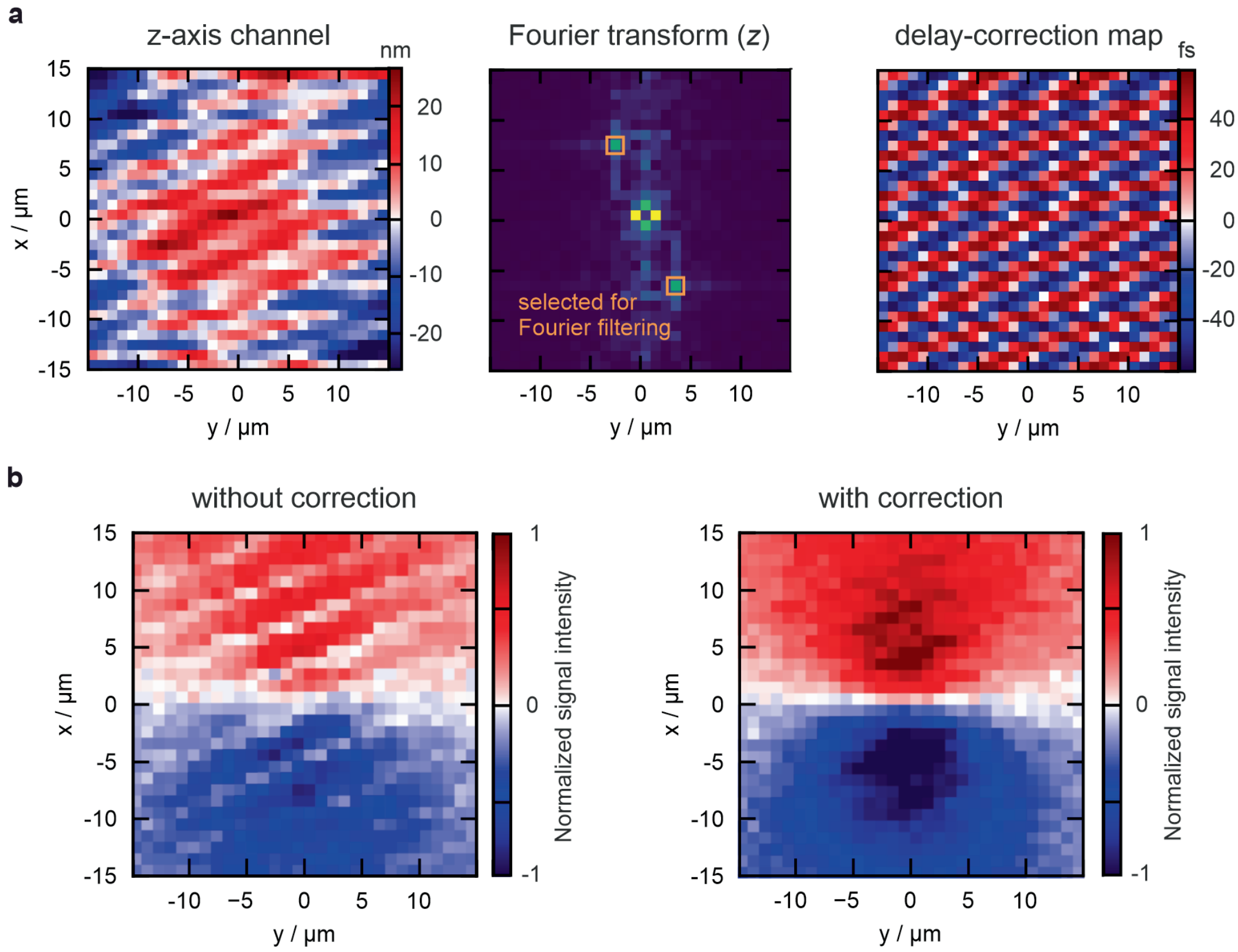

**Fig. S5 | Correction of periodic instruments drifts. a** A delay-correction map is created by applying spatial Fourier filtering to the $z$-channel data and setting the amplitude to 56 fs. **b** Acquired terahertz maps at a fixed delay before and after applying the delay-correction map.

## Acknowledgements

We thank Tobias Kampfrath, Ioachim Pupeza, Maximilian Högner, Jean-François Lampin, and Pierre Koleják for valuable discussions and their general support.

We gratefully acknowledge financial support from the German research foundation (DFG, Deutsche Forschungsgemeinschaft) through the collaborative research centers "SPIN+X" (268565370, TRR 173, project B11) and "Ultrafast spin dynamics" (328545488, TRR 227, project A05), and through the priority program "INTEREST" (468511389, SPP 2413, project ITISA2); from the European Research Council (101088303); from the Chinese Scientific Research Innovation Capability Support Project for Young Faculty (ZYGXQNJSKYCXNLZCXMI3); and from the US National Science Foundation (ECCS-2300152).

## Data availability

The data that support the findings of this study are available at doi.org/10.5281/zenodo.18739933. Additional raw data are available from the corresponding authors upon reasonable request.

## Contributions

FP and RR contributed equally to this work. FP conceived the idea and initiated the project, with facilitation by DM, GvF, DMi, XW, and TSS. FP led the experiments, conducted by FP, JC, MD, FS, and JK. RR provided the theoretical framework and designed and performed all numerical simulations under guidance of TSS. FP processed the measurement data. FP, RR, and TSS interpreted the data. FP, FS, GL, and NT fabricated the samples. FP, RR, and TSS wrote the manuscript and created the figures. All authors contributed to revisions. The project was supervised by DM, NT, GvF, DMi, XW, and TSS.

## References

1 Walowski, J. & Münzenberg, M. Perspective: Ultrafast magnetism and THz spintronics. *Journal of Applied Physics* **120**, doi:10.1063/1.4958846 (2016).

2 Ponseca Jr, C. S., Chábera, P., Uhlig, J., Persson, P. & Sundstrom, V. Ultrafast electron dynamics in solar energy conversion. *Chemical reviews* **117**, 10940-11024 (2017).

3 Carlin, C. C. *et al.* Nanoscale and ultrafast in situ techniques to probe plasmon photocatalysis. *Chemical Physics Reviews* **4** (2023).

4 Kampfrath, T., Kirilyuk, A., Mangin, S., Sharma, S. & Weinelt, M. Ultrafast and terahertz spintronics: Guest editorial. *Applied Physics Letters* **123**, doi:10.1063/5.0167151 (2023).

5 Yao, Z. *et al.* Nanoimaging and nanospectroscopy of polaritons with time resolved s-SNOM. *Advanced Optical Materials* **8**, 1901042 (2020).

6 Müller, M., Paarmann, A. & Ernstorfer, R. Femtosecond electrons probing currents and atomic structure in nanomaterials. *Nature Communications* **5**, 5292 (2014).

7 Zayko, S. *et al.* Ultrafast high-harmonic nanoscopy of magnetization dynamics. *Nature communications* **12**, 6337 (2021).

8 von Korff Schmising, C. *et al.* Imaging ultrafast demagnetization dynamics after a spatially localized optical excitation. *Physical review letters* **112**, 217203 (2014).

9 Guo, X. *et al.* Terahertz nanoscopy: Advances, challenges, and the road ahead. *Applied Physics Reviews* **11** (2024).

10 Koch, M., Mittleman, D. M., Ornik, J. & Castro-Camus, E. Terahertz time-domain spectroscopy. *Nature Reviews Methods Primers* **3**, 48 (2023).

11 Murakami, H. & Tonouchi, M. Laser terahertz emission microscopy. *Comptes Rendus. Physique* **9**, 169-183, doi:10.1016/j.crhy.2007.07.010 (2008).

12 von Hoegen, A. *et al.* Imaging a terahertz superfluid plasmon in a two-dimensional superconductor. *Nature*, doi:10.1038/s41586-025-10082-2 (2026).

13 Ocelic, N., Huber, A. & Hillenbrand, R. Pseudoheterodyne detection for background-free near-field spectroscopy. *Applied Physics Letters* **89**, 101124, doi:10.1063/1.2348781 (2006).

14 Plankl, M. *et al.* Subcycle contact-free nanoscopy of ultrafast interlayer transport in atomically thin heterostructures. *Nature Photonics* **15**, 594-600, doi:10.1038/s41566-021-00813-y (2021).

15 Yao, Z. *et al.* Photo-induced terahertz near-field dynamics of graphene/InAs heterostructures. *Opt Express* **27**, 13611-13623, doi:10.1364/OE.27.013611 (2019).

16 Pizzuto, A., Ma, P. & Mittleman, D. M. Near-field terahertz nonlinear optics with blue light. *Light: Science & Applications* **12**, 96 (2023).

17 Klarskov, P., Kim, H., Colvin, V. L. & Mittleman, D. M. Nanoscale Laser Terahertz Emission Microscopy. *Acs Photonics* **4**, 2676-2680, doi:10.1021/acsphotonics.7b00870 (2017).

18 Pizzuto, A., Mittleman, D. M. & Klarskov, P. Laser THz emission nanoscopy and THz nanoscopy. *Opt Express* **28**, 18778-18789, doi:10.1364/OE.382130 (2020).

19 Li, P. Y. *et al.* Laser terahertz emission microscopy of nanostructured spintronic emitters. *Applied Physics Letters* **120**, 201102 (2022).

20 Cai, J. *et al*. Terahertz spin currents resolved with nanometer spatial resolution. *Applied Physics Reviews* **10** (2023).

21 Paries, F. *et al*. Fiber-tip spintronic terahertz emitters. *Opt Express* **31**, 30884-30893, doi:10.1364/OE.494623 (2023).

22 Seifert, T. *et al*. Efficient metallic spintronic emitters of ultrabroadband terahertz radiation. *Nature Photonics* **10**, 483-488, doi:10.1038/nphoton.2016.91 (2016).

23 Seifert, T. S., Cheng, L., Wei, Z., Kampfrath, T. & Qi, J. Spintronic sources of ultrashort terahertz electromagnetic pulses. *Applied Physics Letters* **120**, 180401 (2022).

24 Rouzegar, R. *et al*. Laser-induced terahertz spin transport in magnetic nanostructures arises from the same force as ultrafast demagnetization. *Physical Review B* **106**, 144427 (2022).

25 Rouzegar, R. *et al*. Femtosecond signatures of optically induced magnons before ultrafast demagnetization. *ArXiv*, doi:10.48550/arxiv.2507.01796 (2026).

26 Hillenbrand, R. & Keilmann, F. Complex optical constants on a subwavelength scale. *Phys Rev Lett* **85**, 3029-3032, doi:10.1103/PhysRevLett.85.3029 (2000).

27 Margetis, D. & Luskin, M. On solutions of Maxwell's equations with dipole sources over a thin conducting film. *Journal of Mathematical Physics* **57** (2016).

28 Niehues, I. *et al*. Nanoscale resolved mapping of the dipole emission of hBN color centers with a scattering-type scanning near-field optical microscope. *Nanophotonics* **14**, 335-342 (2025).

29 Hillenbrand, R., Keilmann, F., Hanarp, P., Sutherland, D. & Aizpurua, J. Coherent imaging of nanoscale plasmon patterns with a carbon nanotube optical probe. *Applied physics letters* **83**, 368-370 (2003).

30 Park, K. D. & Raschke, M. B. Polarization Control with Plasmonic Antenna Tips: A Universal Approach to Optical Nanocrystallography and Vector-Field Imaging. *Nano Lett* **18**, 2912-2917, doi:10.1021/acs.nanolett.8b00108 (2018).

31 Zayats, A. V., Smolyaninov, I. I. & Maradudin, A. A. Nano-optics of surface plasmon polaritons. *Physics reports* **408**, 131-314 (2005).

32 Paries, F. *et al*. Optical damage thresholds of single-mode fiber-tip spintronic terahertz emitters. *Optics express* **32**, 24826-24838 (2024).

33 Lummen, T. T. *et al*. Imaging and controlling plasmonic interference fields at buried interfaces. *Nature communications* **7**, 13156 (2016).

34 Rouzegar, R. *et al*. Broadband Spintronic Terahertz Source with Peak Electric Fields Exceeding 1.5 MV/cm. *Physical Review Applied* **19**, 034018, doi:10.1103/PhysRevApplied.19.034018 (2023).

35 Pizzuto, A. *et al*. Nonlocal time-resolved terahertz spectroscopy in the near field. *ACS photonics* **8**, 2904-2911 (2021).

36 Liao, B. & Najafi, E. Scanning ultrafast electron microscopy: A novel technique to probe photocarrier dynamics with high spatial and temporal resolutions. *Materials Today Physics* **2**, 46-53, doi:10.1016/j.mtphys.2017.07.003 (2017).

37 Li, H. *et al*. Imaging moiré excited states with photocurrent tunnelling microscopy. *Nature materials* **23**, 633-638 (2024).

38 Boschini, F., Zonno, M. & Damascelli, A. Time-resolved ARPES studies of quantum materials. *Reviews of Modern Physics* **96**, 015003 (2024).

39 Šmejkal, L., Sinova, J. & Jungwirth, T. Emerging research landscape of altermagnetism. *Physical Review X* **12**, 040501, doi:10.1103/PhysRevX.12.040501 (2022).

40 Grubišić-Čabo, A. *et al*. Roadmap on quantum magnetic materials. *2D Materials* **12**, 031501 (2025).

41 Müller, M., Martín Sabanés, N., Kampfrath, T. & Wolf, M. Phase-resolved detection of ultrabroadband THz pulses inside a scanning tunneling microscope junction. *ACS photonics* **7**, 2046-2055 (2020).